\documentclass[twocolumn,prb,showpacs]{revtex4}% Physical Review B
\usepackage{graphicx}% Include figure files
\usepackage{dcolumn}% Align table columns on decimal point
\usepackage{bm}% bold math
\usepackage{amsmath}
\usepackage[usenames]{color}

\begin{document}

\title{Angular Dependence of X-ray Absorption Spectrum for Field-aligned\\
       Fe-based Superconductors}
\author{B.\ C.\ Chang}
\author{Y.\ B.\ You}
\author{T.\ J.\ Shiu}
\author{M.\ F.\ Tai}
\author{H.\ C.\ Ku} \email{hcku@phys.nthu.edu.tw}
\affiliation{Department of Physics, National Tsing Hua University,
Hsinchu 30013, Taiwan, Republic of China}
\author{Y.\ Y.\ Hsu}
\affiliation{Department of Physics, National Taiwan Normal
University, Taipei 10677, Taiwan, Republic of China}
\author{L.\ Y.\ Jang}
\author{J.\ F.\ Lee}
\affiliation{National Synchrotron Radiation
Research Center, Hsinchu 30076, Taiwan, Republic of China}
\author{Z.\ Wei}
\author{K.\ Q.\ Ruan}
\author{X.\ G.\ Li}
\affiliation{Hefei National Laboratory for Physical Sciences at
Microscale and Department of Physics, University of Science and
Technology of China, Hefei 230026, China}

\date{\today}

% \pacs{74.70.-b}{Superconducting materials}
% \pacs{74.25.Jb}
% 71.00.00 Electronic structure of bulk materials (see section 73 for
% electronic structure of surfaces, interfaces, low-dimensional
% structures, and nanomaterials; for electronic structure of
% superconductors, see 74.25.Jb)


\begin{abstract}
Anisotropic Fe K-edge and As K-edge X-ray absorption near edge
spectrum (XANES) measurements on superconducting ($T_c$ = 52 K)
(Sm$_{0.95}$La$_{0.05}$)FeAs(O$_{0.85}$F$_{0.15}$) field-aligned
microcrystalline powder are presented. The angular dependence of Fe
pre-edge peak (dipole transition of Fe-$1s$ electrons to
Fe-$3d$/As-$4p$ hybrid bands) relative to the
tetragonal $ab$-plane of aligned powder indicates larger
density of state (DOS) along the $c$-axis,
and is consistent with the LDA band structure calculation.
The anisotropic Fe K-edge spectra exhibit a chemical shift to lower
energy compared to FeO which are closely related to the
itinerant character of Fe$^{2+}$-$3d^6$ orbitals.
The anisotropic As K-edge spectra are more or less
the mirror images of Fe K-edge due to the symmetrical
Fe-As hybridiztion in the FeAs layer.
Angular dependence of As main peak
(dipole transition of As-$1s$ electrons to higher energy
hybrid bands) was observed suggesting character of As-$4d$ $e_g$ orbitals.
\end{abstract}

\maketitle

\section{Introduction}
High-$T_c$ superconductivity with transition temperature $T_c$
up to 55 K was reported in the newly discovered iron-based
RFeAs(O$_{1-x}$F$_{x}$) (rare earth R = La, Ce, Pr, Nd, Sm or Gd)
system.~\cite{p1,p2,p3,p4,p5,p6,p7,p8,p9,p10,p11,p12,p13,p14,p15}
The ZrCuAsSi-type (1111) tetragonal structure (space group P4/nmm)
is a layered structure where the metallic FeAs layers are separated
by the insulating R(O$_{1-x}$F$_{x}$) layers. The discovery of the
iron-based superconductors has generated enormous interest since
these compounds are the first non-cuprate high-$T_c$
superconductors with $T_c$ higher than 50 K. The parent compound
LaFeAsO is semi-metal which shows a spin density wave (SDW) type
antiferromagnetic order below 150 K after the tetragonal to
orthorhombic structural transition.~\cite{p5} Electron doping to the
FeAs layer through partial F$^{-}$ substitution in the O$^{2-}$ site
suppresses both the structural distortion and the magnetic order in
favor of superconductivity.~\cite{p1,p3} The phase diagram of the
RFeAs(O$_{1-x}$F$_{x}$) system (R = La, Ce, Sm) provided a useful
comparison with the phase diagram of the high-$T_c$ cuprate
system.~\cite{p9,p10,p11,p12}

The FeAs layer is believed to be the superconducting layer of the
RFeAs(O$_{1-x}$F$_{x}$) system. Studies on the anisotropic
properties are crucial to understanding this new iron-based system.
Since high-quality single crystals are difficult to grow in the
(1111) system, we use the field-rotation alignment method to align
the superconducting (Sm$_{0.95}$La$_{0.05})$FeAs(O$_{0.85}$F$_{0.15}$)
microcrystalline powder.~\cite{p8} In this report, the anisotropic
Fe K-edge and As K-edge XANES measurements were carried out to
investigate the electronic structure of unoccupied FeAs conduction
bands of the FeAs layer.

\section{experimental}

The superconducting
(Sm$_{0.95}$La$_{0.05}$)FeAs(O$_{0.85}$F$_{0.15}$)
polycrystalline sample was prepared by solid state reaction in an
evacuated and sealed quartz tube. The tetragonal microcrystalline
powder can be aligned at room temperature using the field-rotation
method such that the tetragonal $ab$-plane ($a$ = 0.3936(3) nm) is
parallel to the aligned magnetic field $B_a$ and $c$-axis
($c$ = 0.8495(8) nm) is parallel to the rotation axis.\cite{p8}

Magnetization and magnetic susceptibility data were collected with a
Quantum Design 1-T $\mu$-metal shielded MPMS$_2$ or a 7-T MPMS
superconducting quantum interference device (SQUID) magnetometer
from 2 K to 300 K.

The anisotropic X-ray absorption near edge spectroscopy (XANES) for
the aligned microcrystalline powder was performed at the Taiwan
Light Source (TLS) of the National Synchrotron Radiation Research
Center (NSRRC) in Hsinchu, Taiwan, with Fe K-edge XANES at BL-16A beamline
and As K-edge XANES at BL-17C beamline. Both BL-16A and BL-17C
beamlines use Si(111) double-crystal monochromator (DCM) with the
energy resolution $\Delta E/E = 1.5 \times 10^{-4}$. The tetragonal
$ab$-plane of the aligned sample is placed parallel to the
electric field $\bf{E}$ of the incident linear polarized X-ray.
Fluorescence data was used and Athena program was employed
for background removal and normalization of XANES spectra.~\cite{p16}

\section{results and discussion}

The anisotropic temperature dependence of molar magnetic
susceptibility $\chi_{ab}(T)$ and $\chi_{c}(T)$ for aligned powder
(Sm$_{0.95}$La$_{0.05}$)FeAs(O$_{0.85}$F$_{0.15}$) with applied
field along the $ab$-plane and $c$-axis, respectively, are
shown collectively in Fig.\ 1. For dispersed aligned microcrystalline
in low applied field of 10 G, both zero-field-cooled (ZFC) and
field-cooled (FC) data revealed a sharp superconducting transition
temperature $T_c$ of 52 K, which are identical to the $T_c$
measured from bulk polycrystalline sample. Large ZFC intragrain
Meissner shielding signals were observed with nearly constant
$\chi_{c} = -2.72$ cm$^3$/mol and $\chi_{ab} = -1.13$ cm$^3$/mol
up to 20 K. The anisotropic diamagnetic parameter $\gamma =
\chi_{c}/\chi_{ab}$ of 2.4 was deduced for the aligned
microcrystalline. The FC flux-trapped signal gave the same
anisotropic parameter $\gamma$ of 2.4. Taking into account the
imperfect alignment factor (80\%), a larger anisotropic
diamagnetic parameter $\gamma \sim 3$ was estimated for a perfectly
aligned specimen.\cite{p8}

\begin{figure}
\includegraphics{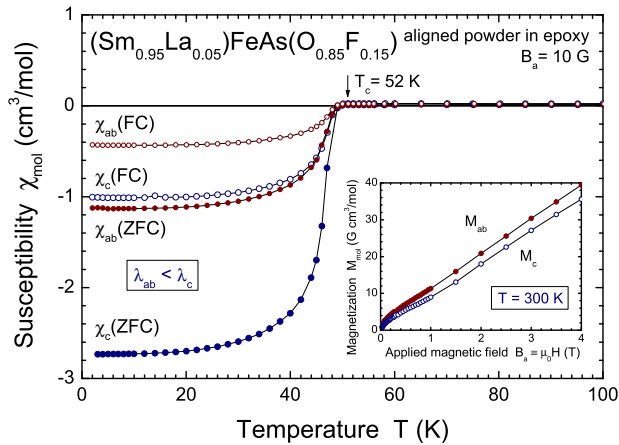}
\caption{\label{label} Anisotropic FC and ZFC
susceptibility $\chi_{ab}(T)$ and $\chi_{c}(T)$ at $B_a$ = 10 G for
(Sm$_{0.95}$La$_{0.05}$)FeAs(O$_{0.85}$F$_{0.15}$) field-aligned powder.
The inset shows the room temperature anisotropic magnetization $M(B_a)$.}
\end{figure}

The anisotropic Fe K-edge spectra for three orientations of the
aligned (Sm$_{0.95}$La$_{0.05}$)FeAs(O$_{0.85}$F$_{0.15}$) powder at
room temperature are shown in Fig.\ 2. The zero angle
($\theta = 0^{\circ}$) refers to that the electric field $\bf{E}$
of the linear polarized X-ray is parallel to the tetragonal
$ab$-plane. Four standards FeO (Fe$^{2+}$), Fe$_{3}$O$_{4}$
(Fe$^{2+/3+}$), Fe$_{2}$O$_{3}$ (Fe$^{3+}$), and Fe metal (with
standard inflection point of 7112 eV) are also presented for
reference.

\begin{figure}
\includegraphics{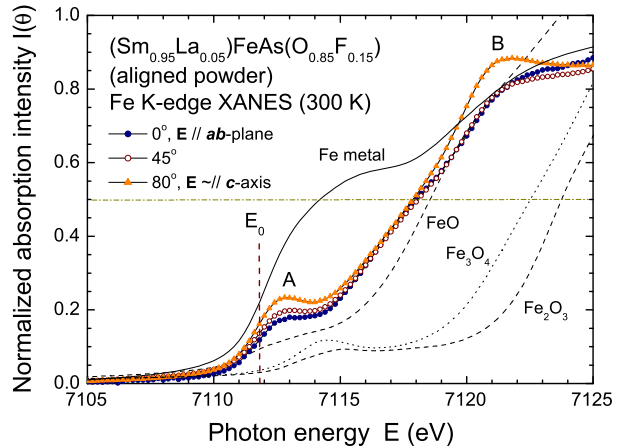}
\caption{\label{label} The anisotropic Fe K-edge XANES spectra for
three orientations of the aligned
(Sm$_{0.95}$La$_{0.05}$)FeAs(O$_{0.85}$F$_{0.15}$) powder. The zero angle
refers to that the electric field $\bf{E}$ of the linear polarized
X-ray is parallel to the tetragonal $ab$-plane.}
\end{figure}

The general feature of the anisotropic Fe K-edge spectra are similar
to the previously reported unoriented random powder data for
isostructural LaFeAs(O$_{1-x}$F$_{x}$) samples.\cite{p13} Two
distinct features of absorption peaks, marked A and B, are observed
in Fig.\ 2. The pre-edge peak A, which is near the inflection point
$E_0$ = 7111.8 eV, relates to the transition of Fe-$1s$ core
electron to the unoccupied Fe-$3d$/As-$4p$ hybrid bands.
The main peak B, which rides on top of the step-onset feature
associated with transitions to continual states, relates to the dipole
transition to Fe-$4p$/As-$4d$ hybrid bands at higher energy. The
energy resolution power $\Delta E$ for Fe K-edge spectra is around 1
eV due to the monochromator used.

The region between the pre-edge peak A and main peak B is referred to as
the main edge. Tracking the chemical shift of the main edge to lower energy
is a standard tool for probing variations of formal valence of Fe ions.
To compare the chemical shift of the superconducting aligned sample with 
four standards, a horizontal dashed line at 50\% of the relative
absorption intensity is drawn in Fig.\ 2. All three anisotropic
(Sm$_{0.95}$La$_{0.05}$)FeAs(O$_{0.85}$F$_{0.15}$) spectra are close
to the FeO (Fe$^{2+}$) standard but are shifted slightly to lower
energy. It is consistent with itinerant character of
Fe$^{2+}$-$3d^6$ orbital in the FeAs conduction layer for the
electron doped (with some Fe$^{1+}$-$3d^7$ character)
(Sm$_{0.95}$La$_{0.05}$)FeAs(O$_{0.85}$F$_{0.15}$) superconductor.

\begin{figure}
\includegraphics{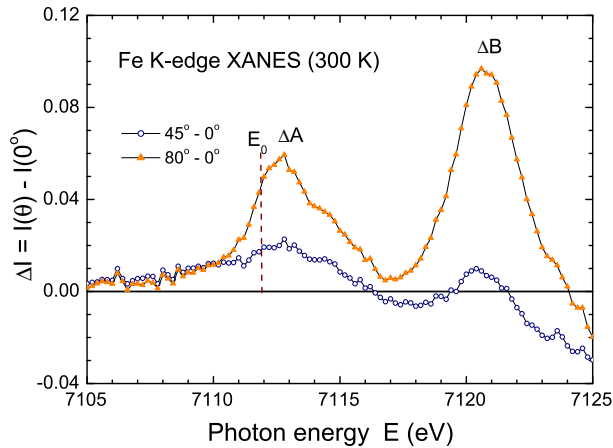}
\caption{\label{label} The difference spectra ($\Delta I$), obtained
by subtracting the $\theta = 0^{\circ}$ data from the other spectra.}
\end{figure}

The pre-edge A spectral feature involve transitions of
Fe-$1s$ core electron into unoccupied final states of Fe-$3d$
either through quadrupole or dipole transitions. Since the Fe $1s$ to
$3d$ quadrupole transition is weak, the observed strong
pre-edge peak A indicates that the local FeAs$_{4}$ tetrahedral ligand
field explicitly allows the dipole transition into unoccupied
Fe-$3d$/As-$4p$ hybrid bands with band width $\sim 3$ eV above the
inflection point $E_0$.

Modest angular dependence of Fe K-edge spectra was observed. For the
anisotropic pre-edge peak A relative to the tetragonal
$ab$-plane of aligned powder, larger density of state (DOS) of
unoccupied Fe-$3d$/As-$4p$ conduction bands close to the
$c$-axis ($\theta = 80^{\circ}$) is observed. This result is
consistent with the local density approximation (LDA) band structure
calculation for this moderately correlated electron system.\cite{p7}
The peak A maximum of 7112.8 eV for $\theta = 80^{\circ}$ and
7113 eV for $\theta = 0^{\circ}$ were observed. Similar
angular dependence was observed in the main peak B with higher DOS
for Fe-$4p$/As-$4d$ hybrid bands along the $c$-axis.

Fig.\ 3 shows the difference spectra ($\Delta I$) which are obtained by
subtracting the $\theta = 0^{\circ}$ spectrum from other Fe K-edge spectra.
Two features labeled $\Delta A$ and $\Delta B$ in the
Fig.\ 3. The large peak $\Delta A$ at $\theta = 80^{\circ}$
indicates a large DOS and large hybridization along the c-axis. A
large peak $\Delta B$ for $\theta = 80^{\circ}$ indicates a large
$1s$-$4p$ absorption edge.

Low temperature anisotropic Fe K-edge measurements, either below or above
$T_c$ = 52 K (at 10 K and 100 K respectively), gives similar pre-edge A
feature as the room temperature data in Fig.\ 4. The DOS of unoccupied
Fe-$3d$/As-$4p$ hybridization above the Fermi level is unaffected
by temperature, which is an indication that the
Fe $3d_{xz}$, $3d_{yz}$ character dominates from normal to
superconducting state. LDA band calculation indicates a Fermi surface
comprised of doubly degenerate hole pocket centered at $\Gamma$
point and a doubly degenerate electron pocket centered at M
point, with dominant Fe $3d_{xz}$, $3d_{yz}$
characters.\cite{p7,p14}

\begin{figure}
\includegraphics{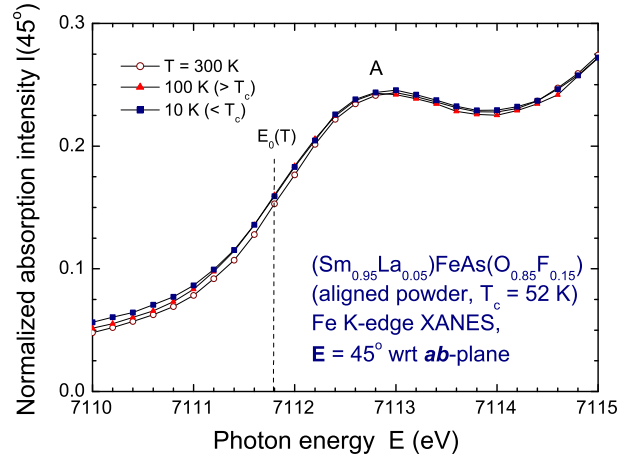}
\caption{\label{label} The  Fe K-edge XANES spectra ($\theta$ =
45$^{\circ}$) for (Sm$_{0.95}$La$_{0.05}$)FeAs(O$_{0.85}$F$_{0.15}$)
at 10, 100, 300 K. Low energy resolution ($\sim$ 1 eV) of
these data fails to resolve the small s-wave superconducting gap
(E$_{g} \sim$ 0.01 eV) below $T_c$.\cite{p6}}
\end{figure}

The anisotropic As K-edge XANES spectra for five orientations of the
aligned (Sm$_{0.95}$La$_{0.05}$)FeAs(O$_{0.85}$F$_{0.15}$) powder at
room temperature are shown collectively in Fig.\ 5. Again, the zero angle
refers to that the electric field $\bf{E}$ of the linear polarized
X-ray is parallel to the tetragonal $ab$-plane. The As
metal standard (with standard inflection point of 11867 eV) is also presented.

\begin{figure}
\includegraphics{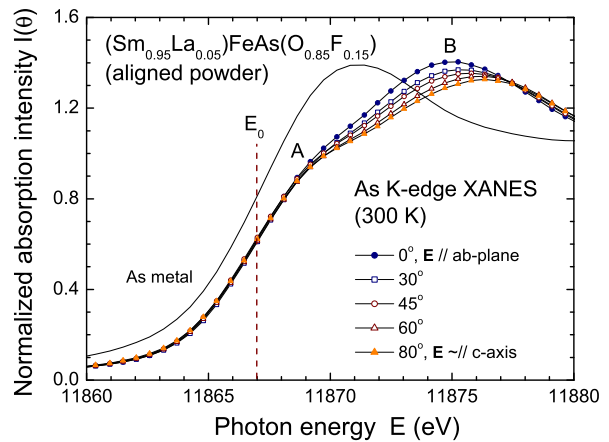}
\caption{\label{label}  The anisotropic As K-edge spectra for
five orientations of the aligned
(Sm$_{0.95}$La$_{0.05}$)FeAs(O$_{0.85}$F$_{0.15}$) powder.}
\end{figure}

The anisotropic As K-edge spectra are more or less the mirror images of
Fe K-edge data due to the Fe-$3d$/As-$4p$ hybridization in
the FeAs conducting layer. The energy resolution power for As K-edge spectra
decreases to $\Delta E \sim$ 2 eV due to higher energy level.
Similar to the Fe K-edge data, two distinct features A and
B are clearly visible in Fig.\ 5. The pre-edge peak A at the onset is related
to the As-$1s$ dipole transition to the unoccupied As-$4p$/Fe-$3d$ hybrid 
bands above the inflection point $E_0$ = 11867 eV.
The main peak B is the transition to higher energy unoccupied
hybrid bands with observable As-$4d$ $e_g$ character.

Weak angular dependence was observed for the pre-edge A feature of
As K-edge spectra. However, stronger angular dependence was seen on main peak B
indicates a larger DOS along the $ab$-planes with peak maximum at 11875 eV
which suggests a hybrid band with As $d_{x^2-y^2}$ characters.
Larger DOS along the c-axis with a peak 
maximum at the slightly higher 11876.5 eV was observed suggesting 
a hybrid band with $d_{3z^2-r^2}$ character.

Fig.\ 6 shows the difference spectra ($\Delta I$) which are obtained
by subtracting the $\theta = 0^{\circ}$ spectrum from other As K-edge spectra.
Two features labeled $\Delta A$ and $\Delta B$ in the
Fig.\ 6. The $\Delta A$ for $\theta  = 45^{\circ}$ has a large DOS
and indicates the hybridization direction of
Fe-$3d$/As-$4p$ bond. The $\Delta B$ decreases with the
orientation angles and indicates the large $1s$-$4p$ absorption
edge of As K-edge spectra along the $ab$-plane.

\begin{figure}
\includegraphics{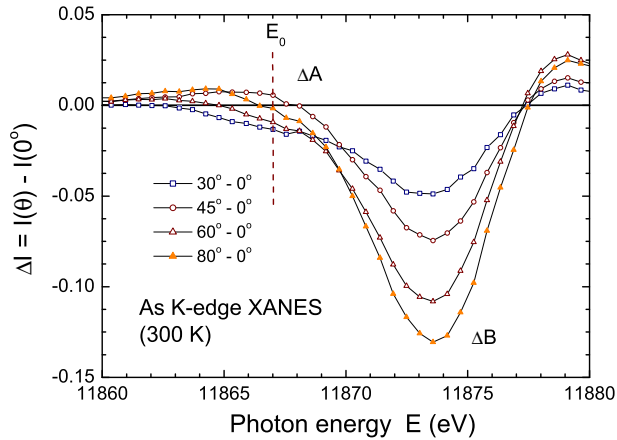}
\caption{\label{label} The difference spectra ($\Delta I$), obtained
by subtracting the zero angle spectrum from the other spectra}
\end{figure}

\section{conclusion}
In conclusion, anisotropic Fe K-edge and As K-edge XANES
measurements on superconducting (Sm$_{0.95}$La$_{0.05}$)FeAs(O$_{0.85}$F$_{0.15}$)
aligned microcrystalline powder are presented. Angular dependence of Fe
pre-edge peak A indicates a larger DOS along the tetragonal $c$-axis, and is
consistent with the LDA band structure calculation. The anisotropic Fe K-edge
spectra exhibits a chemical shift to lower photon energy which is closely related
to the itinerant character of Fe-$3d$ orbitals.
Angular dependence of As main peak B (dipole transition of to higher energy
unoccupied hybrid bands) was observed suggesting character of As-$4d$ $e_g$ orbitals.

\begin{acknowledgments}
This work was supported by NSC95-2112-M-007-056-MY3,
NSC97-2112-M-003-001-MY3, NSC98-2811-M-007-040,
NSFC50421201, MSTC2006CB601003, and MSTC2006CB922005.
\end{acknowledgments}

%\newpage %Just because of unusual number of tables stacked at end

\end{document}